\documentclass[twocolumn]{aastex63}
\usepackage{amsmath}
\usepackage{hyperref}
\usepackage{epstopdf}
\usepackage{breakurl}
\received{2021 July 25}
\revised{2021 August 27}
\accepted{2021 August 30}
\published{2021 November 15}

\begin{document}

\title{Black Hole Gravitational Potential Enhanced Fallback Accretion onto the Nascent Lighter Compact Object: Tentative Evidence in the O3 Run Data of LIGO/Virgo}
\author[0000-0001-9120-7733]{Shao-Peng Tang}
\affil{Key Laboratory of Dark Matter and Space Astronomy, Purple Mountain Observatory, Chinese Academy of Sciences, Nanjing 210033, People's Republic of China}
\affil{School of Astronomy and Space Science, University of Science and Technology of China, Hefei, Anhui 230026, People's Republic of China}
\author[0000-0001-5087-9613]{Yin-Jie Li}
\affil{Key Laboratory of Dark Matter and Space Astronomy, Purple Mountain Observatory, Chinese Academy of Sciences, Nanjing 210033, People's Republic of China}
\affil{School of Astronomy and Space Science, University of Science and Technology of China, Hefei, Anhui 230026, People's Republic of China}
\author[0000-0001-9626-9319]{Yuan-Zhu Wang}
\affil{Key Laboratory of Dark Matter and Space Astronomy, Purple Mountain Observatory, Chinese Academy of Sciences, Nanjing 210033, People's Republic of China}
\author[0000-0002-8966-6911]{Yi-Zhong Fan}
\email{Corresponding author: yzfan@pmo.ac.cn (YZF)}
\affil{Key Laboratory of Dark Matter and Space Astronomy, Purple Mountain Observatory, Chinese Academy of Sciences, Nanjing 210033, People's Republic of China}
\affil{School of Astronomy and Space Science, University of Science and Technology of China, Hefei, Anhui 230026, People's Republic of China}
\author[0000-0002-9758-5476]{Da-Ming Wei}
\affil{Key Laboratory of Dark Matter and Space Astronomy, Purple Mountain Observatory, Chinese Academy of Sciences, Nanjing 210033, People's Republic of China}
\affil{School of Astronomy and Space Science, University of Science and Technology of China, Hefei, Anhui 230026, People's Republic of China}

\begin{abstract}
In a binary system, the gravitational potential of the primary black hole may play an important role in enhancing the fallback accretion onto the lighter compact object newly formed in the second supernova explosion. As a result, the final masses of the binary compact objects would be correlated, as suggested recently by Safarzadeh \& Wysocki. In this work, we analyze the mass distribution of four gravitational-wave events, which are characterized by both a small mass ratio and a low mass ($\leq 5M_\odot$) of the light component, and find tentative evidence for a mass correlation among the objects. To evaluate the feasibility of testing such a hypothesis with upcoming observations, we carry out simulations with a mock population and perform Bayesian hierarchical inference for the mass distribution. We find that with dozens of low mass ratio events, whether there exists correlation in the component mass distributions or not can be robustly tested and the correlation, if it exists, can be well determined.
\end{abstract}

\section{Introduction} \label{sec:intro}
Since the first direct detection of gravitational waves \citep[GWs;][]{2016PhRvL.116f1102A}, two binary neutron star (BNS) and dozens of binary black hole (BBH) merger events have been formally reported \citep{2021PhRvX..11b1053A}. With these observed GW events, the characteristics of the merging BBH population have been extensively investigated in many works, including the observation of some substructures that were not discernible before \citep[e.g.,][]{2021ApJ...913L...7A, 2021arXiv210402969L, 2021ApJ...913L..19T}, the probe of a sharp high-mass cutoff or mass gap in the primary mass distribution \citep[e.g.,][]{2021ApJ...913L...7A, 2021ApJ...916L..16B, 2021ApJ...913L..23E, 2021ApJ...913...42W}, and the examination of relations between component masses \citep{2020ApJ...891L..27F} or correlations between the effective spin and mass ratio \citep{2020ApJ...894..129S, 2021arXiv210600521C}.

The population properties of neutron star\textendash black hole (NSBH) binary systems remain unknown currently, due to the lack of observations (see, e.g., \citealt{2017ApJ...844L..22L} for indirect estimates). Though GW190425 \citep{2020ApJ...892L...3A}, GW190426\_152155 \citep{2021PhRvX..11b1053A}, and GW190814 \citep{2020ApJ...896L..44A} may have probabilities of being NSBH mergers, the nature of both GW190425 and GW190814 relies sensitively on the maximum mass of the NS \citep{2020ApJ...891L...5H, 2021ApJ...908L..28N}, and GW190426\_152155 has a very low signal-to-noise ratio (S/N) that makes its source parameters dominated by the priors \citep{2020arXiv201204978L}. Recently, two confident detections of NSBH events announced by \citet{2021ApJ...915L...5A}, however, change the situation and complete the set of compact binary merger constituents. Previously, though the modeling of the kilonova signal in hybrid GRB 060614 is in favor of an NSBH merger origin \citep{2015ApJ...811L..22J, 2015NatCo...6.7323Y}, such evidence was indirect.

Studying whether the population properties of merging NSBH systems are significantly different from those of BBH or BNS systems is interesting, since different formation channels may leave imprints on population characteristics, with which the underlying astrophysical processes can be better understood. For example, \citet{2020ApJ...892...56T} have investigated the prospect of reconstructing the black hole mass function (BHMF) of NSBH systems and found that the difference (if it really exists) of BHMF between NSBH and BBH systems can be extracted supposing the BBH and NSBH events can be reliably distinguished. Additionally, a common origin for low mass ratio events has been proposed by \citet{2021ApJ...907L..24S}, in which a correlation between component masses should be present. This is because the mass of the primary BH determines the escape velocity of a system, which influences the bound fraction of the ejecta material from the second supernova explosion.

In this work, we focus on the four low mass ratio GW events (i.e., GW190426, GW190814, GW200105, and GW200115). We show that the component-correlated mass scenario is mildly preferred over the component-independent mass distribution, which may support the common origin of these low mass ratio events as hypothesized in \citet{2021ApJ...907L..24S}. We also carry out simulations with a mock population and evaluate the feasibility for probing such correlation with dozens of low mass ratio GW events detected in the full sensitivity run of Advanced LIGO (aLIGO) and Advanced Virgo (AdV). We find that with a reasonable size of $50$ events, the correlated component mass scenario, if it is the case, can be confirmed and the correlation can be well reconstructed. Throughout this work, the uncertainties are for a 68.3\% confidence level unless specifically noted.

\section{Methods}\label{sec:methods}
\subsection{Model of Mass Distributions}\label{sec:model}
We first construct the BHMF with a simple truncated power-law model \citep{2017ApJ...851L..25F, 2019ApJ...882L..24A} that can be described as
\begin{equation}
\begin{aligned} 
&\pi(m_1 \mid \alpha, m_{\rm min}, m_{\rm max}) \\ 
&\propto \begin{cases} (m_1/M_\odot)^{-\alpha} & m_{\rm min} \leq m_1 \leq m_{\rm max} \\ 0 & \mbox{otherwise,} \end{cases}
\end{aligned} 
\end{equation}
where $\alpha$, $m_{\rm min}$, and $m_{\rm max}$ respectively denote the spectral index, low end, and upper end sharp cutoffs of the primary mass ($m_1$) distribution. Meanwhile for the distribution of the secondary mass ($m_2$), a truncated Gaussian model is adopted and a linear dependence for the mean and standard deviation of the Gaussian distribution on $\sqrt{m_1}$ is assumed, i.e.,
\begin{equation}
\begin{aligned} 
&\pi(m_2 \mid m_1, m_{\rm min}, a, \mu_{\rm m}, b, \sigma_{\rm m}) \\ 
&\propto \begin{cases} \exp{\left[-\frac{1}{2}\left(\frac{m_2-\mu_{\rm m}-a\sqrt{m_1}}{\sigma_{\rm m}+b\sqrt{m_1}}\right)^2\right]} & 1M_\odot \leq m_2 \leq m_{\rm min} \\ 0 & \mbox{otherwise.} \end{cases}
\end{aligned} 
\end{equation}
The quantities/parameters associated with mass, including $\{m_1, m_2, m_{\rm min}, m_{\rm max}, \mu_{\rm m}, \sigma_{\rm m}\}$, are in units of solar mass, and the coefficients $a$ and $b$ are in units of $\sqrt{M_\odot}$. Notice that if we set $a=b=0$, it reduces to the component-independent mass distribution model. Alternatively, we also use a uniform distribution for the secondary mass, $\pi(m_2 \mid m_{\rm min})\sim{\rm U}(1M_\odot,m_{\rm min})$, which is found to be consistent with the masses of NSs in GW binaries \citep{2021arXiv210704559L, 2021arXiv210806986L}. Therefore, we have three mass distribution models: the one with a power-law (PL) distribution of primary mass that is correlated with a Gaussian (G) distribution of secondary mass (namely, the ``Correlated PL+G" model), the one identical to the former model but with $a=b=0$ (namely, the ``Independent PL+G'' model), and the one that has a power-law distribution for primary mass but an independent uniform distribution for the secondary mass (namely, the ``Independent PL+U'' model). The mass functions are assumed to be redshift independent, which is likely reasonable because the detection distances of these low mass ratio events are much closer than the coalescing heavy BBH systems. 

\subsection{Mock Population}\label{sec:mock}
\begin{figure*}\setlength{\abovecaptionskip}{-20pt}
    \centering
    \includegraphics[width=0.98\textwidth]{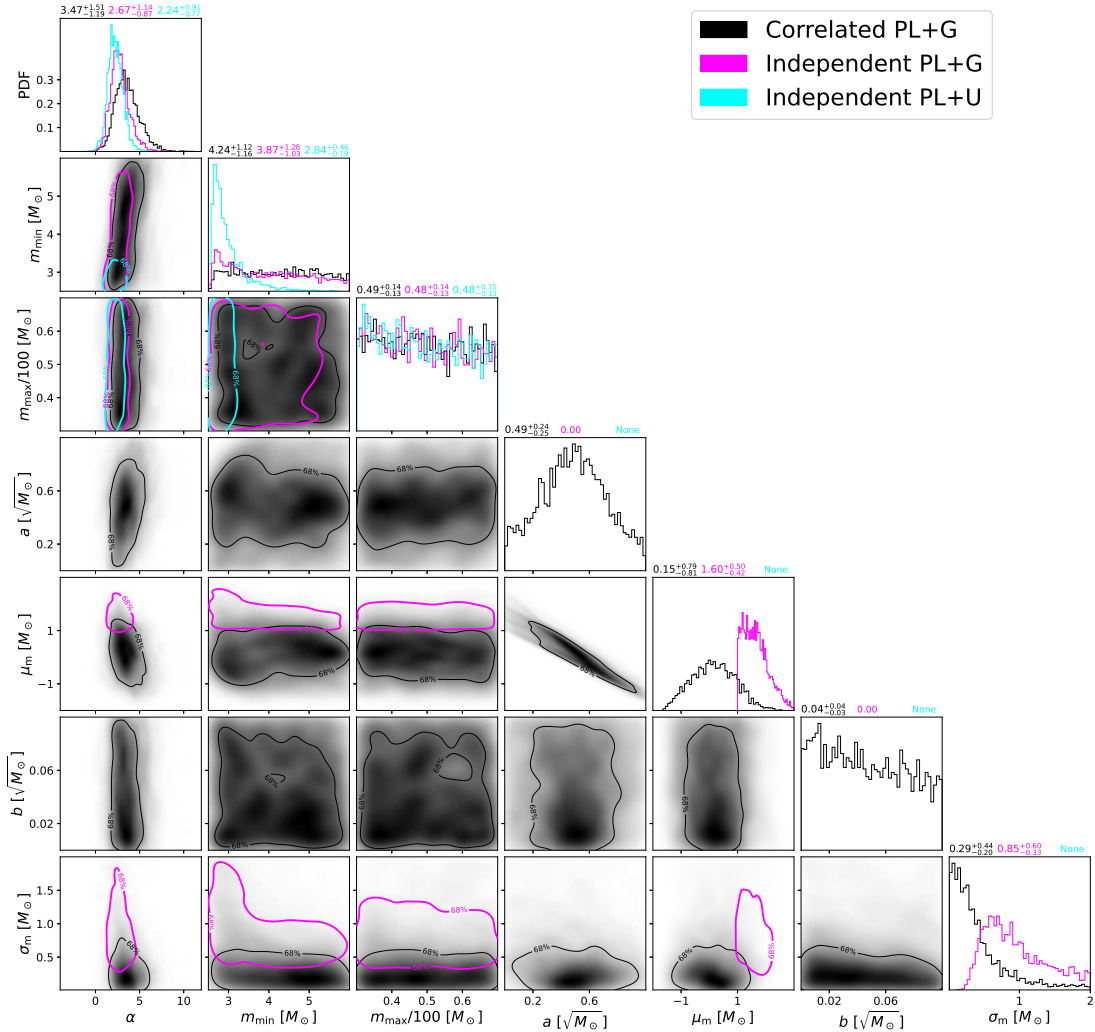}
    \caption{Posterior distributions of the hyperparameters $\Lambda$ inferred with four low mass ratio events. The black, magenta, and cyan lines (including the marginal distributions and contours) represent the Correlated PL+G, Independent PL+G, and Independent PL+U models, respectively. The gray density plots are for the Correlated PL+G model only. Values in the diagonal plots are for the 68.3\% credible intervals.}
    \label{fig:hyperreal}
    \hfill
\end{figure*}
We construct mock populations by setting the fiducial values for parameters $\Lambda=\{\alpha, m_{\rm min}, m_{\rm max}, a, \mu_{\rm m}, b, \sigma_{\rm m}\}$ to $\{2.35, \, 4M_\odot, \, 60M_\odot, \, 0.47\sqrt{M_\odot}, \, 0.37M_\odot, \, 0.05\sqrt{M_\odot}, \\ 0.12M_\odot\}$. To obtain the ``observed'' mass distribution, we inject the simulated signals into the aLIGO and AdV detectors using the {\sc Bilby} package \citep{2019ApJS..241...27A}. For simplicity, we use an aligned spin distribution for the secondary component, i.e., the z-component of spin follows the distribution described in Equation~(A7) of \citet{2018arXiv180510457L} with $\chi_{\rm max}=0.3$ and the spin's tilt angle is taken to be aligned, because binaries born in isolation are expected to form with only small misalignment \citep{2000ApJ...541..319K} and the presence of an accretion disk that has an additional torque may realign the orientation and enhance the magnitude of the spin. We adopt a low (L) spin magnitude distribution with probability density function (PDF) $p(a_{\rm BH})=2(1-a_{\rm BH})$, $a_{\rm BH}\in[0,1)$ as well as an isotropic (I) distribution of the spin's tilt angle with PDF $p(\cos{\theta_{\rm BH}})=1/2\ (-1\leq\cos{\theta_{\rm BH}}\leq1)$ for the BH \citep{2019ApJ...882L..24A} when we perform the injections. The low and isotropic distributions are mainly motivated by the results of \citet{2021ApJ...913L...7A} and the reported negative spin projection onto the orbital angular momentum for GW200115, as well as Figure~3 of \citet{2021ApJ...907L..24S}, where we notice that the effective spin $\chi_{\rm eff}$ is positive and small. Since we only focus on the recovery of the mass distribution model in this work, the assumptions made for spin have just a small influence on our results. The luminosity distance ($d_{\rm L}$) is uniformly distributed in the source frame up to $1500\,{\rm Mpc}$, and the right ascension (R.A.) and declination (decl.) are isotropically distributed. The identification of a GW signal is based on a network S/N threshold $\rho_{\rm th}=12$ and the design sensitivity noise curves\footnote{\url{https://dcc.ligo.org/LIGO-T2000012/public}} \citep{2018LRR....21....3A}.

\subsection{Mock Posteriors}\label{sec:fisher}
After collecting the ``observed'' events, we follow \citet{2019PhRvD.100d3012W} to generate the mock posteriors for each event using an aligned Fisher matrix approach \citep{2008PhRvD..77d2001V, 2013PhRvD..87b4004C, 2014CQGra..31w5009C, 2014PhRvD..89f4048O}. In this approach, the likelihood ($\mathcal{L}_{\rm intrinsic}$) for GW signals is assumed to be Gaussian in three coordinates of $(\mathcal{M}, \eta, \chi_{\rm eff})$, where $\mathcal{M}$, $\eta$, and $\chi_{\rm eff}$ respectively denote the detector-frame chirp mass, the symmetric mass ratio, and the effective spin. We calculate the covariance matrix $\Gamma^{-1}$ of the multivariable Gaussian distribution with the code available in \url{https://git.ligo.org/daniel.wysocki/synthetic-PE-posteriors}, where $\Gamma$ is the approximate Fisher matrix. We set the likelihood mean to a randomly generated value from a multivariable Gaussian distribution with the injected values as mean and covariance matrix $\Gamma^{-1}$. This procedure is to introduce fluctuations and mock the effect of real data analysis, in which slight bias of the parameter inference may be present. Additionally, since Fisher matrix calculation currently works in the detector frame, if we want to obtain the source-frame masses, the inference of luminosity distance should be taken into account. Therefore, we introduce the extrinsic likelihood ($\mathcal{L}_{\rm extrinsic}$) from \citet{2019PhRvD.100h3514C}. For simplicity, we use the {\it Cutler and Flanagan approximation} in \citet{2019PhRvD.100h3514C}, i.e., their Equation~(39) without priors. Thus, our likelihood finally becomes $\mathcal{L}=\mathcal{L}_{\rm intrinsic}\times\mathcal{L}_{\rm extrinsic}$. In practice, we sample the mass ratio $q$ instead of the symmetric mass ratio, and this will not change the likelihood in nest sampling. The posterior samples are generated using the {\sc Bilby} package and {\sc Dynesty} sampler, with uniform in source frame prior on luminosity distance,\footnote{The corresponding redshift is uniform in comoving volume and source-frame time.} isotropic distribution on inclination angle, uniform prior on effective spin $\chi_{\rm eff} \sim {\rm U}(-1, 1)$, and uniform component mass priors (the Jacobians for transforming probability densities are properly considered; see also \citealt{2021arXiv210409508C}) that have been implemented by {\sc Bilby}. The choice of uniform component mass priors is convenient in the hierarchical inference below, where the sampling priors can be easily processed following \citet{2020ApJ...891L..27F}. Using a nonpopulation informed prior of effective spin in the parameter estimation of individual events may lead to biases in the inferred mass distribution as suggested in \citet{2018PhRvD..98h3007N}. Notice that these biases will be important only when we have more than $\sim100$ events; hence, the uniform prior on effective spin may be reasonable and will not produce significant biases for this work.

\subsection{Hierarchical Inference}\label{sec:infer}
With a series of posteriors ($\boldsymbol{d}$) for $N$ events, we introduce the likelihood from \citet{2019PASA...36...10T} to perform the hierarchical inference of hyperparameters $\Lambda$. Regardless of selection effects, the likelihood can be written as
\begin{equation}\label{eq:logllh}
\log[\mathcal{L}(\boldsymbol{d} \mid \Lambda)] \propto \sum_{i}^{N} \log \left[\sum_{k}^{n_i} \frac{\pi(\theta_{i}^k \mid \Lambda)}{\pi(\theta^{k}_{i}\mid{\o})}\right],
\end{equation}
where $N$, $n_{i}$, $\pi(\theta^{k}_{i}\mid\Lambda)$, and $\pi(\theta^{k}_{i}\mid{\o})$ represent the total number of events, the number of downsampled posterior samples, the joint mass distribution, and the prior applied in source parameter inference, respectively.

However, due to the fact that heavier object mergers are relatively easier to detect than the light object mergers, we must take the selection effects into account. Assuming a uniform-in-log prior for merger rate, we can marginalize over the Poisson-distributed rate, and the likelihood in Equation~(\ref{eq:logllh}) can be modified to
\begin{equation}\label{eq:selectlogllh}
\begin{aligned}
\log[\mathcal{L}(\boldsymbol{d} \mid \Lambda)] \propto &\sum_{i}^{N} \log \left[\sum_{k}^{n_i} \frac{\pi(\theta_{i}^k \mid \Lambda)}{\pi(\theta^{k}_{i}\mid{\o})}\right] \\ &-N\times \log[\xi(\Lambda)],
\end{aligned}
\end{equation}
where $\xi(\Lambda)$ represents the detection fraction. We estimate this fraction following the method described in Appendix~A of \citeauthor{2021ApJ...913L...7A} (\citeyear{2021ApJ...913L...7A}; \citealp[see also][]{2018CQGra..35n5009T, 2019RNAAS...3...66F, 2020arXiv200705579V}). To accelerate the injection campaign, we first filter signals with the network S/N$\leq 10$ (such a value is estimated analytically following \citealt{2019PhRvD.100h3514C}; see also \citealt{1993PhRvD..47.2198F, 1994PhRvD..49.2658C}), otherwise we inject them to the detectors and numerically calculate the S/N. The number of effective injections is $\sim10^5$, which is sufficiently large for our hierarchical inference \citep{2019RNAAS...3...66F}. Then the detection fraction can be obtained using weighted Monte Carlo integration over found injections ($\rho>\rho_{\rm th}=12$). And we use the hierarchical inference module of the {\sc Bilby} package \citep{2019ApJS..241...27A} to infer the hyperparameters with the {\sc Dynesty} sampler \citep{2020MNRAS.493.3132S}. Finally, for the priors of the hyperparameters, we use uniform distributions for parameters $\alpha$, $m_{\rm min}$, $m_{\rm max}$, $a$, $\mu_{\rm m}$, $b$, and $\sigma_{\rm m}$ with the ranges of $[-4,12]$, $[2.5,6]\,M_\odot$, $[30,70]\,M_\odot$, $[0,1]\,\sqrt{M_\odot}$, $[-2, 3]\,M_\odot$, $[0,0.1]\,\sqrt{M_\odot}$, and $[0.01,2]\,M_\odot$, respectively.

\section{Results} \label{sec:results}
\begin{figure*}
    \centering
    \includegraphics[width=0.98\textwidth]{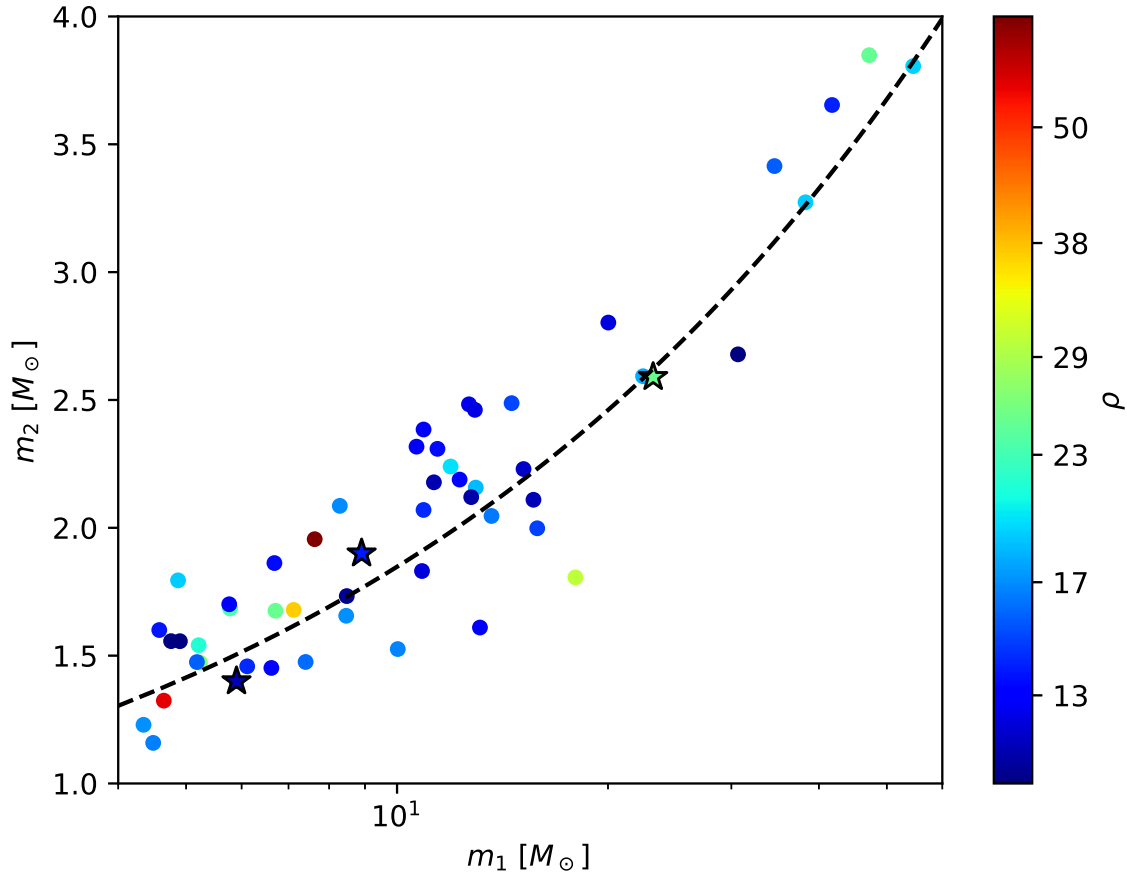}
    \caption{Source-frame component masses of ``observed'' samples collected from the injections of mock population with the fiducial values mentioned in Section~\ref{sec:mock}. Points with a star marker represent the measurements of real events, i.e., GW190814, GW200105, and GW200115. Different colors encode the corresponding network S/N. The dashed line tracing the correlation between primary and secondary mass is just for illustration.}
    \label{fig:injections}
    \hfill
\end{figure*}
First, we use the posterior samples\footnote{Download from \url{https://dcc.ligo.org/LIGO-P2000223/public}, \url{https://dcc.ligo.org/public/0175/P2100143/002/GW200105_162426_posterior_samples_v2.h5}, and \url{https://dcc.ligo.org/public/0175/P2100143/002/GW200115_042309_posterior_samples_v2.h5}.} of four low mass ratio GW events (i.e., GW190426, GW190814, GW200105, and GW200115) to perform hierarchical inference for the parameters that characterize the mass distribution. For GW190426 and GW190814, the `PublicationSamples' in the HDF5 files are used, while for GW200105 and GW200115, the `C01:Combined\_PHM\_low\_spin' samples are used. Since the individual-event posteriors were calculated under priors that are flat in detector-frame masses, we reassign event-level priors for source-frame masses following the method described in \citet{2020ApJ...891L..27F}. And the actual O3 noise curves used to perform injection campaign are obtained from \url{https://dcc.ligo.org/LIGO-T2000012/public} \citep{2018LRR....21....3A}. The results of three models constructed in Section~\ref{sec:model} are presented in Figure~\ref{fig:hyperreal}.
\begin{figure*}\setlength{\abovecaptionskip}{-20pt}
    \centering
    \includegraphics[width=0.98\textwidth]{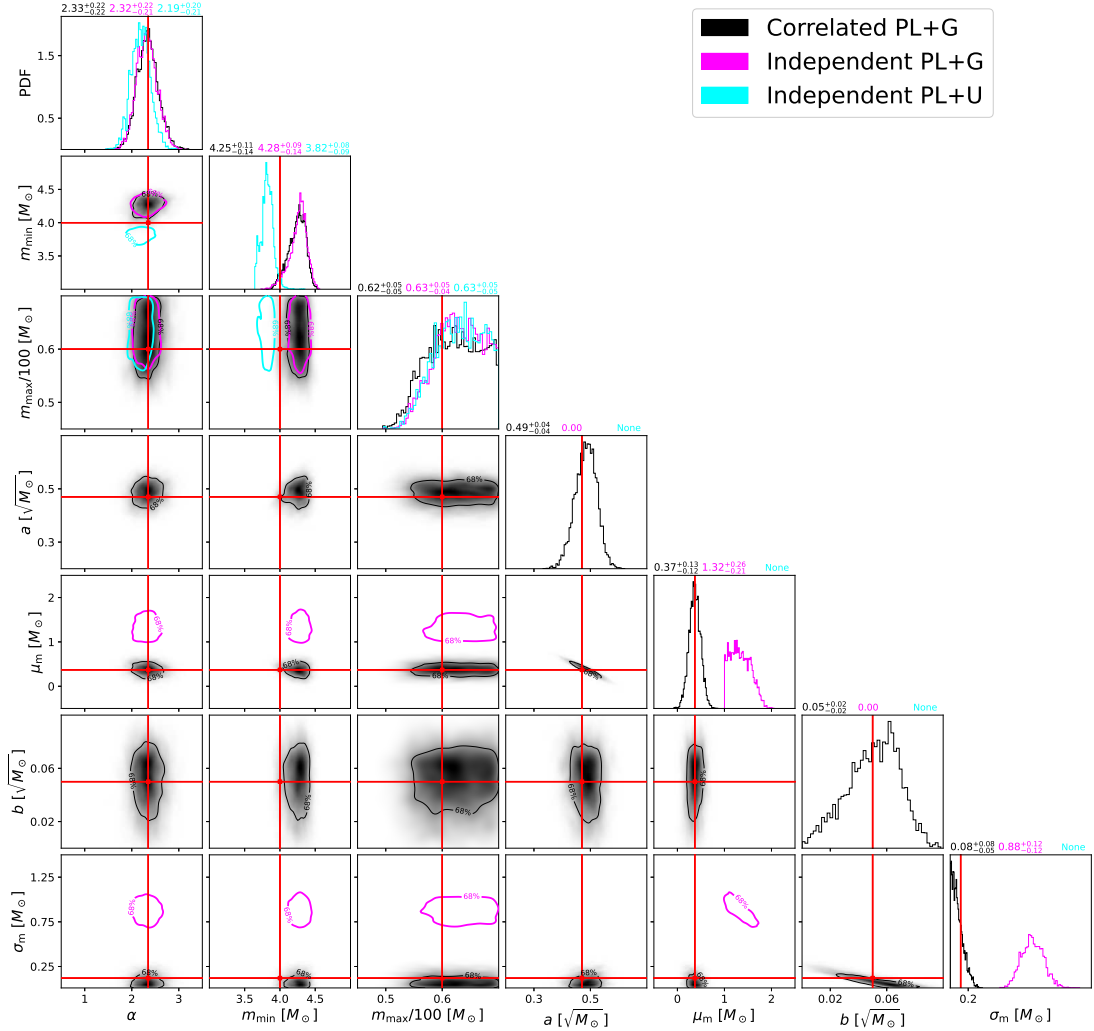}
    \caption{Same as Figure~\ref{fig:hyperreal}, but inferred with mock ``observation'' data. Red solid lines denote the fiducial values used to generate the mock population. With about $50$ events the presence of a correlation among the component masses can be reliably probed.}
    \label{fig:hypersim}
    \hfill
\end{figure*}
In the case of the Correlated PL+G model, we notice that the coefficients $a$ and $\mu_{\rm m}$, which determine the mean of the secondary mass Gaussian distribution, are constrained to $0.49_{-0.25}^{+0.24}\,\sqrt{M_\odot}$ and $0.15_{-0.81}^{+0.79}\,M_\odot$, respectively. Meanwhile, we find that a strong correlation is present in their posterior distributions. The power-law index ($\alpha$) of the BHMF is constrained to $3.47_{-1.19}^{+1.51}$, while other parameters remain unconstrained, which is an expected consequence of inferring mass distribution with only four events. In the case of the Independent PL+G model, the index $\alpha$ and the mean $\mu_{\rm m}$ are constrained to $2.67_{-0.87}^{+1.14}$ and $1.60_{-0.42}^{+0.50}\,M_\odot$, respectively. We also notice that if we use a narrow prior range for the standard deviation $\sigma_{\rm m}$ for this model, e.g., $[0.01, 0.6]$, $\sigma_{\rm m}$ will converge to the upper bound. This is because the secondary mass of GW190814, which is significantly heavier than those of the other events, becomes an ``outlier'' for the Independent PL+G model, and a Gaussian distribution with narrow width is difficult to cover such a measurement. While the Correlated PL+G model can shift $\mu_{\rm m}$ according to the primary mass, this allows the Gaussian distribution to cover a wider range in secondary mass compared with the Independent PL+G model. In the case of the Independent PL+U model, the index $\alpha$ is constrained to $2.24_{-0.77}^{+0.91}$, and different from the previous two models, the $m_{\rm min}$ is peaked at $2.84\,M_\odot$, relying on the most massive secondary mass we used. The logarithm Bayes factor between the Correlated PL+G and Independent PL+G models is evaluated to $\ln \mathcal{B}=0.8$, showing that current observations mildly favor the correlated mass distribution. Meanwhile, the evidence of Independent PL+G and Independent PL+U models is comparable, with $\ln \mathcal{B}=0.2$. We also adopt the Akaike information criterion \citep[AIC;][]{Hirotugu1981Likelihood}, ${\rm AIC}=2n-\log \mathcal{L}_{\rm max}$, to perform model comparison. We find that $\Delta{\rm AIC}=-6.8$ for the Correlated PL+G model versus the Independent PL+G model and $\Delta{\rm AIC}=3.7$ for the Independent PL+G model versus the Independent PL+U model, which means that the Correlated PL+G model is still the most preferred, while the Independent PL+U model becomes more favored than the Independent PL+G model.

It is quite interesting to investigate whether we can reliably extract information about the correlation from future GW observations if the dependence really holds. Here we evaluate the prospect of probing such a correlation using an artificially constructed mock population. We collect a number of $50$ ``observed'' events shown in Figure~\ref{fig:injections}, and assign posteriors for the mass measurements for each event using the method described in Section~\ref{sec:fisher}. Then similar to the real data analysis, we perform hierarchical inference for the hyperparameters. We find that with a reasonable size of $50$ low mass ratio events, the power-law index ($\alpha$) and $m_{\rm min}$ in the BHMF can be well reconstructed (as shown in Figure~\ref{fig:hypersim}), which is consistent with the results of \citet{2020ApJ...892...56T}. More intriguingly, the coefficients $a$ and $\mu_{\rm m}$ are also well recovered with small uncertainties, and the logarithm Bayes factor between the Correlated PL+G and Independent PL+G models is evaluated to $\ln \mathcal{B}=38.3$, while between the Independent PL+G and Independent PL+U models, $\ln \mathcal{B}$ is evaluated to $24.0$. We therefore conclude that with future GW detections, the correlation between the component masses can be robustly probed and the parameters can be well determined.

\section{Summary and Discussion} \label{sec:discussion}
In this work, we analyze the mass distribution of four low mass ratio events within three models, i.e., the Correlated PL+G, Independent PL+G, and Independent PL+U models. We find that these events mildly prefer the component-correlated mass distribution, with a relation of $m_2\simeq0.5\sqrt{m_1/M_\odot}+0.25\,[M_\odot]~(m_1\gtrsim3.5\,M_\odot)$, where $m_1$ ($m_2$) is the primary (secondary) mass. We then evaluate the feasibility for probing such a correlation with future observations. By generating the mock ``observations'' from an artificially constructed population and using them to perform Bayesian hierarchical inference, we find that with a reasonable size of $50$ low mass ratio events, whether there exists correlation or not can be tested and the correlation (if it exists) can be well determined. Assuming that the O4 run will last for $1.5$yr and the reachable ranges for O4 and O5 are $\sim1.5$ and $\sim2.5$ times that of O3 \citep{2018LRR....21....3A}, we can roughly estimate $T\sim(50-1.5\times4\times1.5^3)/(4\times2.5^3)\sim0.5$yr, which means that the number of events, $50$, may be reached after a half year of the beginning of O5 run. Our results indicate that the common origin model proposed by \citet{2021ApJ...907L..24S} is promising for explaining current observational data, and may be corroborated by upcoming GW detections. We know that NS mass distributions in double NS systems and NS\textendash white dwarf (NSWD) systems are different \citep{2013ApJ...778...66K}, since the NS in NSWD will accrete material from its companion. Meanwhile for an NS in NSBH systems, if the newly born NS has an initial mass similar to that of double NS systems, through accretion of the bounded ejecta material \citep{2021ApJ...907L..24S}, the NS evolved from such scenario is also expected to have a shifted mass distribution, which can be checked by future observations. We do not incorporate the information from component spins, which could also carry imprints from the accretion, and we will jointly analyze both the mass and spin population properties in our future works. The mock posteriors used in this work are an ideal case, and the effective Fisher matrix approach may not completely represent the situation of real data analysis, since biases in parameter estimates could be complicated. However, with more and more loud events accumulated in the future observing runs, a few biased events will not change the overall population properties.

\acknowledgments
We thank the anonymous referee for the helpful suggestions. This work was supported in part by NSFC under grants No. 11921003, No. 11933010, and No. 12073080, as well as the Chinese Academy of Sciences via the Strategic Priority Research Program (grant No. XDB23040000) and the Key Research Program of Frontier Sciences (No. QYZDJ-SSW-SYS024).

\software{Bilby \citep[version 1.1.2;][\url{https://git.ligo.org/lscsoft/bilby/}]{2019ApJS..241...27A}, Dynesty \citep[version 1.0.1;][\url{https://dynesty.readthedocs.io/en/latest/}]{2020MNRAS.493.3132S}, PyCBC \citep[version 1.16.11;][\url{https://github.com/gwastro/pycbc}]{2020zndo....596388N}}

\bibliography{ms.bib}{}
\bibliographystyle{aasjournal}

\end{document}